\documentclass[twocolumn,prd,nofootinbib,aps,prl,floats,floatfix,superscriptaddress,amsmath,amssymb,longbibliography,secnumarabic,preprintnumbers]{revtex4-1} %
\usepackage{graphicx}% Include figure files
\usepackage{dcolumn}% Align table columns on decimal point
\usepackage[utf8]{inputenc}

\makeatletter
\def\thesection{\arabic{section}}%
\def\p@section{}%
\def\thesubsection{\thesection.\arabic{subsection}}%
\def\p@subsection{}%
\def\thesubsubsection{\thesubsection.\arabic{subsubsection}}%
\def\p@subsubsection{}%
\def\appendix{%
    \par
    \setcounter{section}\z@
    \setcounter{subsection}\z@
    \setcounter{subsubsection}\z@
    \def\thesubsection{\thesection.\arabic{subsection}}%
    \def\thesubsubsection{\thesubsection.\arabic{subsubsection}}%
    \def\p@subsection{}%
    \def\p@subsubsection{}%
    \@addtoreset{equation}{section}%
    \def\theequation@prefix{\thesection}%
    \addtocontents{toc}{\protect\appendix}%
    \@ifstar{%
    \def\thesection{\unskip}%
    \def\theequation@prefix{A.}%
    }{%
    \def\thesection{\Alph{section}}%
    }%
}%
\makeatother

\usepackage{bm}% bold math
\usepackage{hyperref}
\usepackage{xcolor}
\usepackage{float}
\usepackage{siunitx}
\usepackage[capitalise]{cleveref}
\usepackage[sort&compress]{natbib}
\usepackage{booktabs}

\begin{document}

\setlength{\abovedisplayskip}{5pt}
\setlength{\belowdisplayskip}{5pt}
\setlength{\abovedisplayshortskip}{5pt}
\setlength{\belowdisplayshortskip}{5pt}

\newcommand{\jk}[1]{\textcolor{orange}{ #1}}
\newcommand{\pc}[1]{\textcolor{blue}{ \textbf{PC: #1}}}
\newcommand{\ZT}[1]{\textcolor{teal}{ \textbf{ZT: #1}}}

\newcommand{\changed}[1]{{\color{red} \textbf{#1}}}

\preprint{IFT-UAM/CSIC-24-164}
% \begin{flushright}
%     IFT-UAM/CSIC-24-164
% \end{flushright}

\title{Clarity through the Neutrino Fog: Constraining New Forces in Dark Matter Detectors}

\author{Pablo Blanco-Mas}
\email{pablo.blanco@ift.csic.es}
\affiliation{Instituto de Física Teórica UAM-CSIC, Calle Nicolás Cabrera 13–15, Universidad Autónoma de Madrid, 28049 Madrid, Spain}

\author{Pilar Coloma}
\email{pilar.coloma@ift.csic.es}
\affiliation{Instituto de Física Teórica UAM-CSIC, Calle Nicolás Cabrera 13–15, Universidad Autónoma de Madrid, 28049 Madrid, Spain}

\author{Gonzalo Herrera}
\email{gonzaloherrera@vt.edu}
\affiliation{Center for Neutrino Physics, Department of Physics, Virginia Tech, Blacksburg, VA 24061, USA}

\author{Patrick Huber}
\email{pahuber@vt.edu}
\affiliation{Center for Neutrino Physics, Department of Physics, Virginia Tech, Blacksburg, VA 24061, USA}

\author{Joachim Kopp}
\email{jkopp@cern.ch}
\affiliation{Theoretical Physics Department, CERN, 1 Esplanade des Particules, CH-1211 Geneva 23, Switzerland}
\affiliation{PRISMA+ Cluster of Excellence $\&$ Mainz Institute for Theoretical Physics, 55128 Mainz, Germany}

\author{Ian M. Shoemaker}
\email{shoemaker@vt.edu}
\affiliation{Center for Neutrino Physics, Department of Physics, Virginia Tech, Blacksburg, VA 24061, USA}

\author{Zahra Tabrizi}
\email{z$\_$tabrizi@pitt.edu}
\affiliation{Theoretical Physics Department, CERN, 1 Esplanade des Particules, CH-1211 Geneva 23, Switzerland}
\affiliation{PITT PACC, Department of Physics and Astronomy, University of Pittsburgh, 3941 O’Hara St., Pittsburgh, PA 15260, USA}

\begin{abstract}
The PandaX-4T and XENONnT experiments present indications of Coherent Elastic Neutrino Nucleus Scattering (CE$\nu$NS) from ${}^{8}$B solar neutrinos at 2.6$\sigma$ and 2.7$\sigma$, respectively. This constitutes the first observation of the neutrino ``floor" or ``fog", an irreducible background that future dark matter searches in terrestrial detectors will have to contend with. Here, we first discuss the contributions from neutrino--electron scattering and from the Migdal effect in the region of interest of these experiments, and we argue that they are non-negligible. Second, we make use of the recent PandaX-4T and XENONnT data to derive novel constraints on light scalar and vector mediators coupling to neutrinos and quarks. We demonstrate that these experiments already provide world-leading laboratory constraints on new light mediators in some regions of parameter space.
\end{abstract}

%=============================================================================
\maketitle
%=============================================================================

%=============================================================================
\section{Introduction}
%=============================================================================

Recently, the PandaX-4T collaboration has reported the first measurement of the neutrino fog from coherent elastic neutrino--nucleus scattering (CE$\nu$NS) of solar ${}^{8}$B neutrinos \cite{PandaX:2024muv} at $2.6\sigma$. This was soon followed by an announcement of the XENONnT collaboration, which reported the measurement of CE$\nu$NS at $2.7\sigma$ \cite{XENON:2024ijk}. Both experiments observed the signal using the combined ionization and scintillation (S1+S2) signal. PandaX-4T also reported a signal in the S2 (ionization) channel separately. Interestingly, while for the combined analysis both experiments report background expectations consistent with the observed event rates, this is not the case for the S2-only analysis at PandaX-4T, where an excess with respect to the background expectation of a few tens of events is observed. Using a Poissonian likelihood of the number of signal and background events, the excess has a low significance of $1.3\sigma$ for science Run0, but a somewhat larger significance of $3.6\sigma$ for Run1. Thus, being conservative, for PandaX-4T we will use only the Run0 data.

The fact that dark matter direct detection experiments could provide sensitivity to Beyond the Standard Model (BSM) interactions of solar neutrinos was pointed out about a decade ago, and a series of works have followed through the years discussing various BSM scenarios, different interaction channels of neutrinos with the detector, and the complementarity of different experiments and detector technologies \cite{Pospelov:2011ha, Harnik:2012ni, Cerdeno:2016sfi, Dutta:2017nht, AristizabalSierra:2019ykk, Shoemaker:2020kji, Herrera:2023xun, Amaral:2023tbs, DeRomeri:2024dbv, Xia:2024ytb, Majumdar:2024dms}. In light of the recent datasets from XENONnT and PandaX-4T, some works have constrained non-standard interactions of neutrinos with quarks \cite{Li:2024iij, AristizabalSierra:2024nwf},\footnote{The XENONnT and PandaX-4T datasets have also been used to constrain the uncertainty in the weak mixing angle at low momentum transfer in Ref.~\cite{Maity:2024aji}.} corresponding to new mediators with masses much larger than the momentum transfer of the scattering process, of order $q = \sqrt{2 m_A E_{\rm nr}} \lesssim \SI{10}{MeV}$ at direct detection experiments. (Here, $m_A$ is the mass of the target nucleus and $E_{\rm nr}$ is the nuclear recoil energy.) Here we will study instead the current sensitivity of PandaX-4T and XENONnT to \emph{light mediators} via CE$\nu$NS of ${}^{8}$B solar neutrinos at the detector, making use of the combined (S1+S2) signal as well as the PandaX-4T S2-only dataset. We will consider in particular light vector mediators with universal couplings to all quarks and neutrinos, as well as light scalar mediators with either universal couplings, or couplings proportional to the quark masses (as in the case of coupling through mixing with the SM Higgs field). We will demonstrate that in some regions of parameter space, these experiments provide complementary and even stronger constraints to other experiments sensitive to CE$\nu$NS such as COHERENT \cite{COHERENT:2017ipa}, whose sensitivity to light new mediators has been widely studied in a series of works, \textit{e.g} \cite{Denton:2018xmq, AristizabalSierra:2018eqm, Abdullah:2018ykz, Khan:2019cvi, Miranda:2020tif, Banerjee:2021laz, AtzoriCorona:2022moj}.\footnote{We will restrict ourselves to derive upper limits on the coupling of the mediator to quarks and neutrinos, even though the excess events in the S2-channel of PandaX-4T could in principle be interpreted as a hint for a positive signal.}

The paper is organized as follows: In \cref{sec:formalism}, we discuss the different scattering processes that neutrinos can undergo in a liquid xenon detector, showing that CE$\nu$NS yields the dominant event rate in the Region of Interest (ROI) of the PandaX-4T and XENONnT experiments, although the Migdal effect and elastic neutrino--electron scattering can yield sizable contributions (\cref{sec:interactions-sm}. We then discuss how extensions of the SM with light scalar or vector mediators can modify these rates (\cref{sec:interactions-scalar,sec:interactions-vector}. We present and discuss our results in \cref{sec:results}, showing parameter space exclusions for the different BSM scenarios. We conclude in \cref{sec:conclusions}.

%=============================================================================
\section{Neutrino interactions in liquid xenon detectors}
\label{sec:formalism}
%=============================================================================

Solar neutrinos reaching direct detection experiments on Earth can leave ionization signatures in liquid noble gas detector through three distinct processes: CE$\nu$NS \cite{Billard:2013qya, OHare:2021utq}, the Migdal effect following CE$\nu$NS \cite{Bell:2019egg, Herrera:2023xun}, and elastic neutrino--electron scattering \cite{Essig:2018tss, Wyenberg:2018eyv, Carew:2023qrj}. In this section, we will discuss the event rates for all these processes. We will argue that in the Regions of Interest (ROIs) of the recent PandaX-4T and XENONnT analyses, CE$\nu$NS is indeed the dominant process, but the Migdal effect and neutrino--electron scattering can also yield sizable contributions.

%-----------------------------------------------------------------------------
\subsection{Standard Model}
\label{sec:interactions-sm}
%-----------------------------------------------------------------------------

\subsubsection{CE$\nu$NS}
%------------------------

In the SM, the differential cross section for CE$\nu$NS, as a function of the nuclear recoil energy $E_{\rm nr}$, reads \cite{Freedman:1973yd}
\begin{align}
    \frac{d\sigma_{\nu N-\nu N}}{dE_{\rm nr}}
        = \frac{G_F^2 m_A}{\pi} Q_V^2 \bigg(1 - \frac{m_A E_{\rm nr}}{2 E_\nu^2}\bigg) \,,
    \label{eq:xsec-sm-7s}
\end{align}
where $m_A$ is the mass of the target nucleus and $Q_V$ is its vector charge,
\begin{align}
    Q_V = (g_{p V} Z + g_{n V} N) \, F_A(E_{\rm nr}) \,,
\end{align}
with $F_N$ denoting the nuclear form factor. Here we adopt the prescription from \cite{ENGEL1991114, Harnik:2012ni}. Here, $Z$ and $N$ stand for the numbers of protons and neutrons in the nucleus, respectively, and the SM neutral current vector couplings are $g_{p V} = \frac{1}{2} - 2\sin ^2 \theta_W$ and $g_{n V} = -\frac{1}{2}$, with $\theta_W$ the Weinberg angle. The differential scattering rate is obtained from the convolution of the differential cross section with the solar neutrino flux 
\begin{align}
    \frac{dR^{\rm CE\nu NS}}{dE_{\rm nr}} = \epsilon(E_{\rm nr}) \, N_T \int_{E_\nu^{\min}}^{E_\nu^{\max}} \!
        \frac{d\Phi}{dE_\nu} \frac{d\sigma_{\nu N \to \nu N}}{dE_{\rm nr}} dE_\nu
\end{align}
where $N_T$ the number of xenon (Xe) target nuclei in the detector, $\epsilon(E_{\rm nr})$ is the detection efficiency of the experiment, and $\frac{d\Phi}{dE_\nu}$ is the differential neutrino flux as a function of the neutrino energy.

In the limit where the nuclear mass largely exceeds the nuclear recoil energy, $m_A \gg E_{\rm nr}$ (applicable here since we only consider recoil energies as large as \SI{3}{keV}), the minimum neutrino energy needed to induce a recoil $E_{\rm nr}$ is given by
\begin{align}
    E_{\nu}^{\rm min} = \sqrt{m_A E_{\rm nr} / 2} \, ,
    \label{eq:E-nu-min-nr}
\end{align}
while $E_{\nu}^{\rm max} $ is the kinetic endpoint for the $^8\mathrm{B}$ spectrum.  We take the expected $^8\mathrm{B}$ flux from \cite{1963ApJ...137..344B,Vinyoles:2016djt, Vitagliano:2019yzm}, consistent with measurements from Borexino \cite{Borexino:2017uhp}. For CE$\nu$NS, we have checked that in nuclear recoil energy windows of the PandaX-4T and XENONnT analyses, the contribution from ${}^{8}$B largely exceeds the contribution from the other solar neutrino fluxes. For neutrino--electron scattering, on the other hand, it is the contribution from $pp$ solar neutrinos \cite{Vitagliano:2019yzm} that dominates.

\subsubsection{Migdal Effect}
%----------------------------

Part of the ionization (S2) signal in a nuclear recoil event is due to the Migdal effect \cite{migdal:1939svj}, which describes the response of atomic electrons to the nucleus receiving a kick. The corresponding differential cross section is obtained by multiplying the cross section for neutrino--nucleus scattering with an ionization form factor $|Z_{\rm ion}(E_{\rm er})|^2$, which accounts for the ionization probabilities of the xenon atom orbitals, see Ref.~\cite{Herrera:2023xun}. This form factor depends on the electron recoiling energy $E_{\rm er}$ and is given by
\begin{align}
    |Z_{\rm ion}(E_{\rm er})|^2 = \frac{1}{2\pi} \sum_{n, l} p_{n l \to E_{\rm er}} \,,
\end{align}
where the differential transition probability of an electron in the orbital $(n, l)$ to an unbound state with recoil energy $E_{\rm er}$ is denoted by $p_{n l \to E_{\rm er}}$. We take the ionization probabilities for Xe from Ref.~\cite{Ibe:2017yqa}, taking into account the most external orbitals $5p$, $5s$, $4d$, $4p$, $4s$, $3d$, $3p$, and $3s$ (the number of Migdal events for electrons in the internal orbitals will be subleading and can be safely ignored). The electron equivalent energy spectrum due to the Migdal effect can then be calculated as
\begin{align}
    \frac{dR^{\rm mig}}{dE_{\rm det}} &= N_T \int dE_{\rm nr} \, dE_{\rm er} \,
        \delta(E_{\rm det} - q_{\rm nr} E_{\rm nr} - E_{\rm er} + |E^{nl}|) \nonumber \\
    &\times  \epsilon(E_{\rm nr}) \int_{E_\nu^{\min}}^{E_\nu^{\max}} \! dE_\nu \,
        \frac{d\Phi}{dE_\nu} \frac{d\sigma_{\nu N \to \nu N}}{dE_{\rm nr}} \times |Z_{\rm ion}(E_{\rm er})|^2 \,,
\end{align}
where $E_{\rm det}$ is the measured electron equivalent energy in a liquid Xe detector. Here, $E_{\rm nr}$ is the nuclear recoil energy, $E_{\rm er}$ is the recoil energy of the ionized electron, $E^{nl}$ is its orbital binding energy before the interaction, and $q_{\rm nr}$ is the quenching factor that relates nuclear recoil energies to electron--equivalent recoil energies, $E_\mathrm{er} = q_\mathrm{nr}(E_\mathrm{nr})E_\mathrm{nr}$. The integration over nuclear recoils lies in the kinematical range
\begin{equation}
\frac{\left(E_{\rm er}+|E^{nl}|\right)^2}{2 m_A}<E_{\rm nr}<\frac{\left(2 E_\nu-\left(E_{\rm er}+|E^{nl}|\right)\right)^2}{2\left(m_A+2 E_\nu\right)}.
\end{equation} 

As previously discussed, we take $q_{\rm nr} = 0.15$~\cite{Ibe:2017yqa}. It should be noted that the quenching factor (or scintillation efficiency) in liquid xenon is uncertain for low-energy nuclear recoils, with different sets of data reporting measurements in the range $q_{\rm nr} = 0.05$ to $0.2$, see, \textit{e.g.}, \cite{Plante:2011hw, Mu:2013dga, Szydagis_2011, XENON100:2013smi}. In this regard, an interesting observation is that the measurements obtained in Refs.~\cite{PandaX:2024muv, XENON:2024ijk}, interpreted in the context of the solar neutrino flux in the Standard Model (SM) and without any BSM contributions to the interaction cross section, disfavor values of $q_{nr}$ at low as 0.02 (albeit with a low significance).

\subsubsection{Neutrino--Electron Scattering}
%--------------------------------------------

Neutrinos can also scatter off electrons in the atom elastically. In this case, the cross section can be calculated using the free-electron approximation, which accounts for the binding energy of electrons in the atom via a Heaviside function \cite{Hsieh:2019hug,Coloma:2022avw, Herrera:2024ysj}.\footnote{A calculation including the atomic wave functions of xenon yields an event rate smaller by 20--25\% \cite{Chen:2016eab}.} In this approximation, the differential scattering cross section for neutrinos of flavor $\alpha = e, \mu, \tau$ on electrons is
\begin{align}
    \frac{d\sigma_\alpha}{dE_{\rm er}} =
        \sum_{n,l} \theta(E_{\rm er} - |E^{nl}|)
        \frac{d\sigma^{0}_\alpha}{dE_{\rm er}} ,
\end{align}
with the cross section for neutrino scattering on free electrons ($\nu_\alpha + e \to \nu_\alpha + e)$ given by \cite{Vogel:1989iv}
\begin{align}
    \frac{d\sigma^{0}_\alpha}{dE_{er}} = \frac{2 G_F^2 m_e}{\pi}
        \bigg[g_{L\alpha}^2 \!
            + g_{R\alpha}^2 \bigg(\! 1-\frac{E_{\rm er}}{E_\nu} \!\bigg)^2 \!
            - g_{L\alpha} g_{R\alpha} \frac{m_e E_{\rm er}}{E_\nu^2}\bigg] .
    \label{eq:xsec-sm-nu-e}
\end{align}
Here, $g_{Le} =  (g_V + g_A)/2 + 1$, $g_{L\mu} = g_{L\tau} = (g_V + g_A)/2$, and $g_{R\alpha}  = (g_V - g_A)/2$, with $g_V = -1/2 + 2 \sin^2\theta_W$ and $g_A = -1/2$. The differential recoil rate is then
\begin{align}
    \frac{dR}{dE_{\rm er}} = \epsilon N_T \sum_\alpha \!\!
        \int_{E_\nu^{\min}}^{E_\nu^{\max}} \!\! dE_\nu \frac{d\phi^\alpha}{dE_\nu}
        \frac{d\sigma_\alpha}{dE_{\rm er}}
\label{eq:dRdE-sm}
\end{align}
where the minimum neutrino energy required to induce a given recoil energy is now given by 
\begin{align}
    E_\nu^{\min} = \frac{E_{\rm er}
                 + \sqrt{2 m_e E_{\rm er} + E_{\rm er}^2}}{2} \,.
\end{align}
(Note that, in contrast to \cref{eq:E-nu-min-nr}, we have here kept terms suppressed by $E_{\rm er} / m_e$, given that this ratio, while still small, is larger than the corresponding one for nuclear recoils, $E_{\rm nr} / m_A$.)

\begin{figure}[H]
    \centering
    \includegraphics[width=1\linewidth]{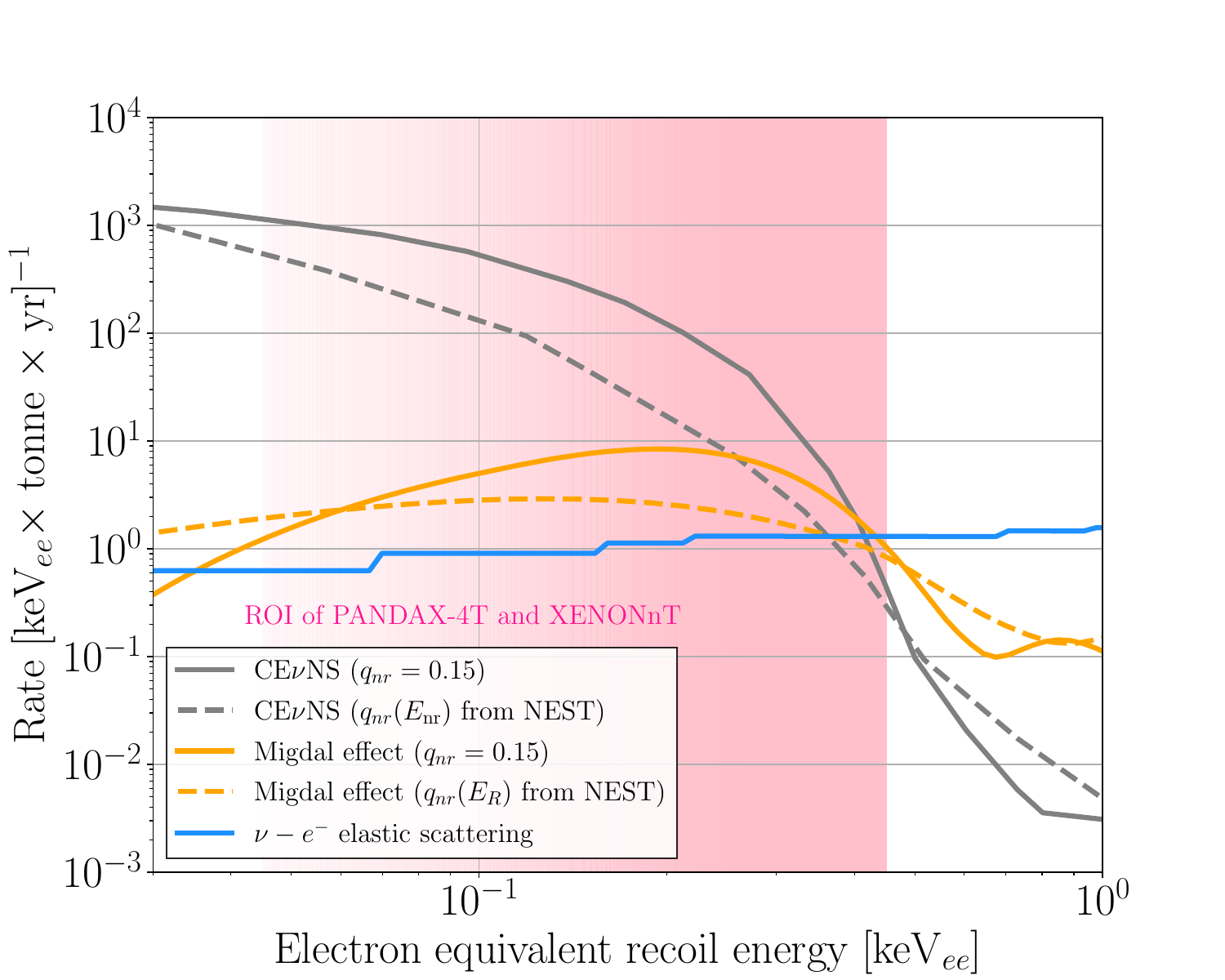}
    \caption{Comparison of ionization rates induced by ${}^{8}$B solar neutrinos in a liquid xenon detector via different SM processes: CE$\nu$NS (grey), the Migdal effect (orange), and elastic neutrino--electron scattering (blue). For CE$\nu$NS, we consider a quenching factor of $q_{nr} = 0.15$ (solid lines) and the quenching factor from NEST (dashed lines) \cite{Szydagis_2011}, to translate nuclear recoil energies into electron-equivalent recoil energies. For comparison, we show in shaded pink the approximate ROI of the recent PandaX-4T and XENONnT analyses of solar neutrinos \cite{PandaX:2024muv, XENON:2024ijk}. The color gradient highlights the fact that the efficiency function is not constant at the ROI of these experiments, but rather increases with energy. The signal induced by CE$\nu$NS clearly dominates in the ROI of these experiments, but the Migdal and neutrino--electron scattering contributions are non-negligible. Note that here we have not applied any efficiency factors to the expected ionization rates.}
    \label{fig:rate}
\end{figure}

The ionization rates due to CE$\nu$NS, the Migdal effect, and neutrino--electron scattering are shown in \cref{fig:rate}, for two different choices of the quenching factor as indicated.
%%%%%%%%%%%%%%%%%%%
%%%%%%%%%%%%%%%%%%%%%%
We see that CE$\nu$NS is the dominant contribution in the low recoil energy ROI from Refs.~\cite{PandaX:2024muv, XENON:2024ijk}, while the contributions from the Migdal effect and from elastic neutrino--electron scattering are smaller, but not entirely negligible. In particular, they are comparable to CE$\nu$NS at the upper end of the ROI, and at higher energies they dominate. This raises the question to what extent the Migdal effect and neutrino--electron scattering may affect the PandaX-4T and XENONnT results, which do not explicitly include these contributions in their signal predictions \cite{PandaX:2024muv, XENON:2024ijk}. For the combined S1+S2 analyses, neglecting neutrino--electron scattering is certainly justified as these events yield a larger S2/S1 ratio than CE$\nu$NS, allowing for efficient discrimination. This is also true for the Migdal effect involving $K$-shell and $L$-shell electrons \cite{Xu:2023wev, bang2023migdal}. $M$- and $N$-shell electrons, however, which are ejected from their host atoms more frequently than inner-shell electrons due to their lower binding energies, lead to signals very similar to the main nuclear recoil signal. They may therefore not be negligible.

In an analysis using only the S2 signal both neutrino--electron scattering and the Migdal effect contribute. Still, for the PandaX-4T S2-only analysis discussed here, the corresponding contributions are small. Using the efficiency function from Ref.~\cite{PandaX:2024muv}, the total number of neutrino--electron scattering events in the ROI is predicted to be $O(0.1)$, while Migdal effect contributes $O(1)$ event. This should be compared to the total number of 43 expected solar neutrino CE$\nu$NS events in the ROI.

The relative importance of the Migdal effect and neutrino--electron scattering compared to CE$\nu$NS depends crucially on the quenching factor (or scintillation efficiency) of liquid xenon at such small energies, which is uncertain. The qualitative discussion presented previously should therefore be taken with a grain of salt. The values that we use, either $q_{\rm nr} = 0.15$ or the quenching factor model from NEST~\cite{Szydagis_2011}, are in overall good agreement with the majority of measurements of the quenching factor \cite{Plante:2011hw, Mu:2013dga, Szydagis_2011, XENON100:2013smi} and with the Lindhard model prediction \cite{osti_4701226} at recoil energies above $E_{\rm nr} \gtrsim 3$ keV, but uncertainties remain large at lower recoil energies.

In our search for physics beyond the SM, we will only include CE$\nu$NS and neutrino--electron scattering, neglecting the Migdal effect. This approach can be regarded conservative since including additional contributions to the SM background would reduce the room there is for new physics. \footnote{In general, physics beyond the SM will also contribute to the Migdal effect. However, as long as the new physics-induced event rates are much smaller than the SM ones, and the Migdal/CE$\nu$NS and $\nu$--$e$ scattering/CE$\nu$NS ratios are similar for new physics processes and SM processes, it remains true that neglecting the Migdal effect and neutrino--electron scattering is conservative.} We also stress that our CE$\nu$NS analysis is performed using nuclear recoil energies, as this is how the collaborations provide their efficiencies. In other words, our results rely on the same assumptions made by the collaborations regarding the quenching factor.

%-----------------------------------------------------------------------------
\subsection{A New Scalar Mediator}
\label{sec:interactions-scalar}
%-----------------------------------------------------------------------------

In extensions of the SM neutrinos may experience new interactions with nuclei. We will consider the possibility that neutrinos couple to a new scalar mediator $\phi$ which also couples to quarks. Then, we will then extend the discussion also to new vector mediators. For the scalar case, the relevant terms in the Lagrangian read
\begin{align}
    \mathcal{L} \supset \phi \left( g_\nu \bar{\nu}_R\nu_L \,
                      + \, g_q \bar{q} q + \mathrm{h.c.} \right)
                      - \frac{1}{2}m_\phi^2 \phi^2 \,,
    \label{eq:L-scalar}
\end{align}
where $m_\phi$ is the mass of the scalar and $g_q$, $g_\nu$ are dimensionless coupling constants. We will consider both the case that $g_q$ is the same for all quark flavors and the case that $g_q$ scales with the quark masses. The latter scenario is motivated by models in which the new scalar couples to quarks through mixing with the SM Higgs boson. It corresponds to the replacement
\begin{align}
    g_q \to \frac{g_q m_q}{v_H}
    \label{eq:q-mass-dependence}
\end{align}
in \cref{eq:L-scalar}, where $v_H = \SI{246}{GeV}$ is the SM Higgs vev. In the absence of left-right neutrino mixing, the contribution due to a new scalar interaction would not interfere with the SM one because of the different Lorentz structures\footnote{While left-right mixing arises naturally in neutrino mass models, its value depends completely on the mechanism responsible for neutrino masses. For this reason we have decided not to include an interference term for a scalar mediator in our analysis. }. The scalar-induced CE$\nu$NS cross section reads~\cite{Cerdeno:2016sfi, Farzan:2018gtr}
\begin{align}
    \bigg( \frac{d \sigma_{\nu N-\nu N}}{dE_{\rm nr}} \bigg)_\phi
        = \frac{ g_{\nu}^2 \mathcal{Q}_\phi^2 m_A^2}{4 \pi}
          \frac{E_{\rm nr}}{E_\nu^2 (2 m_A E_{\rm nr}+m_\phi^2)^2} ,
    \label{eq:xsec-scalar}
\end{align}
where the scalar ``charge'' is defined as \cite{Bertuzzo:2017tuf}
\begin{align}
    \mathcal{Q}_\phi = \bigg[ Z \sum_{q=u,d,s} g_q \frac{m_p}{m_q} f_{T_q}^p
                  + N \sum_{q=u,d,s} g_q \frac{m_n}{m_q} f_{T_q}^n \bigg] F_N(E_{\rm nr}) .
\end{align}
We take values of the hadronic form factors $f_{T_q}^p$ and $f_{T_q}^n$ from Ref.~\cite{Hoferichter:2015dsa}. We have compared our definition of the scalar charge for Xenon with that from \cite{Hoferichter_2020}, which is computed using a nuclear shell model, finding that at the low recoil energies of interest to us, the two expressions differ by at most 10$\%$.

The scalar-mediated scattering rate is then obtained in full analogy to \cref{eq:dRdE-sm}, but with the SM cross-section replaced by \cref{eq:xsec-scalar}.

%-----------------------------------------------------------------------------
\subsection{A Leptophilic Scalar}
\label{sec:leptophilic}
%-----------------------------------------------------------------------------

While we are primarily interested in neutrino scattering on nuclei in this paper, we will also investigate a scenario where the scattering is predominantly on electrons. In particular, let us consider a leptophilic light scalar mediator, whose Lagrangian reads 
\begin{align}
    \mathcal{L}_{\phi,e} \supset \phi \left( g_{\nu} \bar{\nu}_R \nu_L
                            + g_{e} \bar{e} e + \mathrm{h.c.} \right)
                      - \frac{1}{2}m_\phi^2 \phi^2 .
    \label{eq:L-leptophilic}
\end{align}
The corresponding elastic neutrino--electron scattering cross section reads (for a neutrino with flavor $\alpha$)
\begin{align}
    \frac{d\sigma^\phi_\alpha}{dE_{\rm er}}
        = \frac{g_\nu^2 g_e^2 m_e^2 E_{\rm er}}
               {4\pi E_\nu^2 \big( 2 m_e E_{\rm er} + m_\phi^2 \big)^2}.
    \label{eq:xsec-leptophilic}
\end{align}

%-----------------------------------------------------------------------------
\subsection{A New Vector Mediator}
\label{sec:interactions-vector}
%-----------------------------------------------------------------------------

We now consider a new light vector mediator $Z'$ coupling universally to all quarks and neutrinos, thus interacting at the detector predominantly via CE$\nu$NS. The vector mediator may also induce scattering off electrons via kinetic mixing with the SM photon \cite{Holdom:1985ag}. Constraints on such new mediators can therefore be derived also from electron scattering experiments, see \textit{e.g.}\ Refs.~\cite{Khan:2020csx, DeRomeri:2024dbv}, leading to the conclusion that kinetic mixing needs to be very small, given current theoretical and experimental bounds \cite{Caputo:2021eaa, Hebecker:2023qwl, Cline:2024wja}. We will therefore work with the simplified Lagrangian (neglecting kinetic mixing)
\begin{align}
    \mathcal{L} \supset g_q \bar{q} \gamma^\alpha Z_\alpha^\prime q
                      + g_\nu \bar{\nu} \gamma^\alpha Z_\alpha^\prime \nu \,.
    \label{eq:L-vector}
\end{align}
Note that, unlike for the scalar mediator, the $Z'$-mediated CE$\nu$NS amplitude can interfere with the SM one. The total scattering cross section taking both contributions into account is obtained from \cref{eq:xsec-sm-7s} by replacing the vector charge according to
\begin{align}
    Q_V \to Q_V + \frac{3 g_q g_\nu}{\sqrt{2} G_F}
            \frac{(Z + N) \, F_N(E_{\rm nr})}{2 m_A E_{\rm nr} + m_{Z^{\prime}}^2} \,.
    \label{eq:QV-vector}
\end{align}
Again the differential event rate then follows from \cref{eq:dRdE-sm}. %In the following, we assume positive couplings $g_{q}g_{\nu}$, which leads to constructive interference in the scattering cross section.
Since the weak charge is negative (being dominated by the SM coupling to neutrons), a positive product of the new couplings $g_\nu g_q$ will lead to a destructive interference. 

%=============================================================================
\section{PandaX-4T and XENONnT Constraints on Light Mediators}
\label{sec:results}
%=============================================================================

%-----------------------------------------------------------------------------
\subsection{Main Results}
%-----------------------------------------------------------------------------

\begin{figure}
    \centering
    \includegraphics[width=\columnwidth]{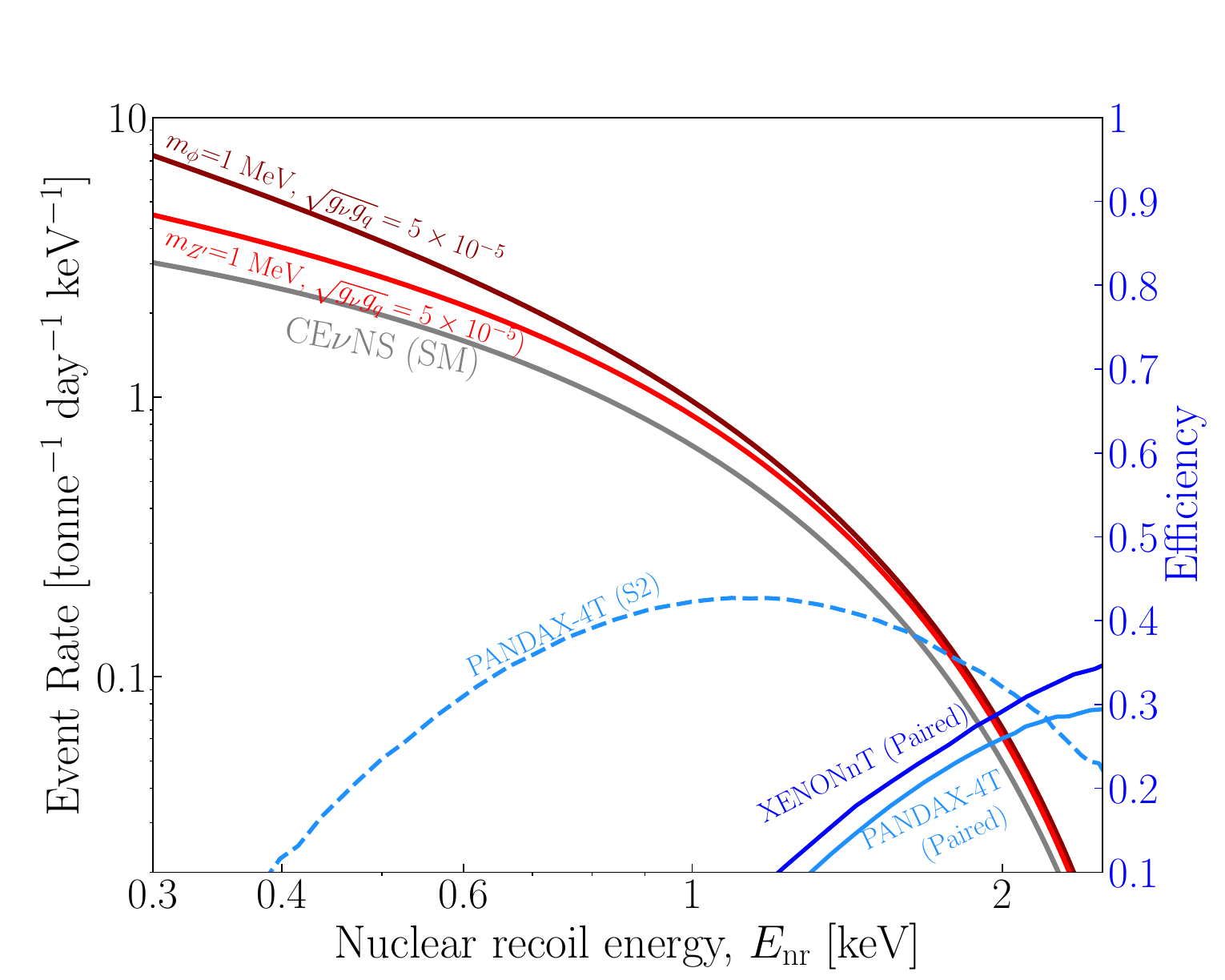}
    \caption{Scattering rates for CE$\nu$NS in the SM (gray), compared to scenarios with a new flavor-universal light scalar (dark red) or vector (light red) mediator. In both cases, we have assumed flavor-universal couplings. We also show the total efficiencies of the PandaX-4T and XENONnT analyses from Refs.~\cite{PandaX:2024muv, XENON:2024ijk} (blue curves / right vertical axis). The plot clearly reveals the enhancement $\propto 1/E_{\rm nr}$ of the scalar-induced scattering processes compared to the SM at low energies, as expected from \cref{eq:xsec-scalar}.}
    \label{fig:eff}
\end{figure}

We are now ready to calculate the expected neutrino scattering event rates in PandaX-4T and XENONnT, and to set limits on possible new physics contributions to the observed CE$\nu$NS rate. We begin in \cref{fig:eff} by showing the differential scattering rate for the SM and for extensions featuring a scalar or vector mediator, both with flavor-universal couplings to quarks. In both BSM scenarios, we observe an increased scattering rate compared to the SM at low energies. This can be easily understood from \cref{eq:xsec-scalar,eq:QV-vector}: as long as $m_\phi$ is negligible compared to the typical momentum transfer in the scattering process, the differential cross section scales as $1/E_{\rm nr}$. Considering the detection efficiencies shown in blue in \cref{fig:eff}, we see that liquid xenon detectors are beginning to be sensitive to recoil energies low enough to benefit from this enhancement. Nevertheless, detection of nuclear recoil energies below \SI{1}{keV} remains challenging. We note the appreciable difference between the efficiency curve for the S2-only analysis in PandaX-4T compared to XENONnT: PandaX-4T is sensitive down to lower recoil energies, while XENONnT's maximum efficiency is higher, though only at nuclear recoil energies above \SI{1}{keV}.

\begin{table}
    \centering
    \begin{ruledtabular}
    \begin{tabular}{lllll}
                                          & \multicolumn{2}{c}{PandaX-4T} & \multicolumn{1}{c}{XENONnT} \\
                                          & S1+S2 & S2 only               & S1+S2 \\%& S2 only \\
         \hline
         exposure [tonne yrs]             & 1.20  & 1.04                  & 3.51\\  %& 3.51    \\
         observed events $N_{\rm obs} $   & 1     & 158                   & 37\\    %& 133     \\
         background events $N_{\rm bck} $ & 2.16  & 144                   & 38.3\\  %& 135.9
         \hline
         signal events scalar benchmark   & 1.04    &  30.52                   & 6.55\\
         signal events vector benchmark   & 1.11    &  44.32                   & 8.05\\
    \end{tabular}
    \end{ruledtabular}
    \caption{Experimental parameters used in our analysis, based on Ref.~\cite{PandaX:2024muv} for PandaX-4T  and on Ref.~\cite{XENON:2024ijk} for XENONnT. The last two rows show numbers of expected signal events for two benchmark scenarios, corresponding, respectively, to a new flavor-universal scalar or vector with for $m_{\phi, Z'} = \SI{1}{MeV}$, $g_\phi = 5.2\times 10^{-6}$, $g_{Z'} = 1.5\times 10^{-5}$.Note that here the background is defined including the SM contribution to the solar neutrino event rates.  }
    \label{tab:exp-params}
\end{table}

\begin{figure*}
    \centering
    \begin{tabular}{cc}
        \includegraphics[width=0.48\textwidth]{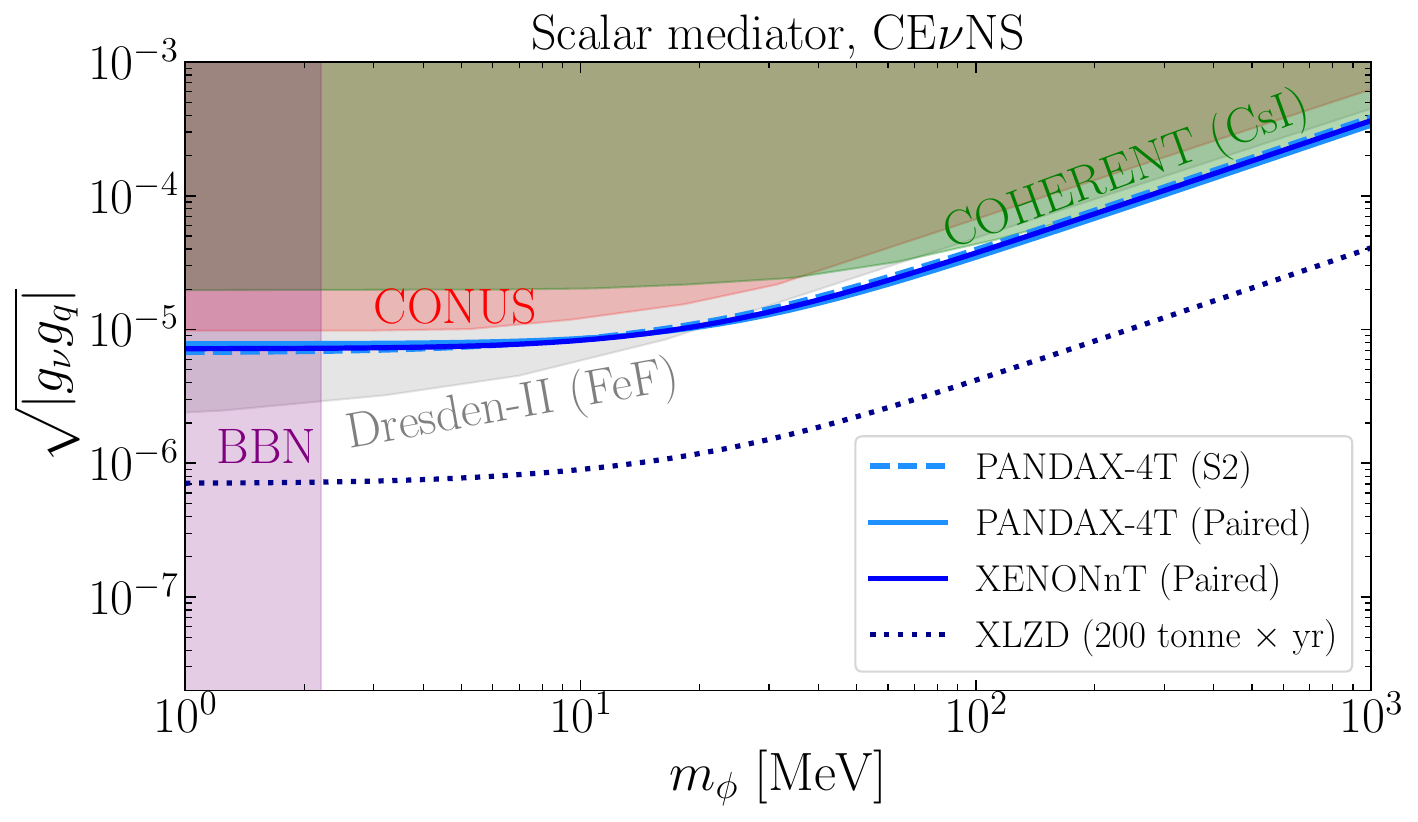} \hfill &
        \includegraphics[width=0.48\textwidth]{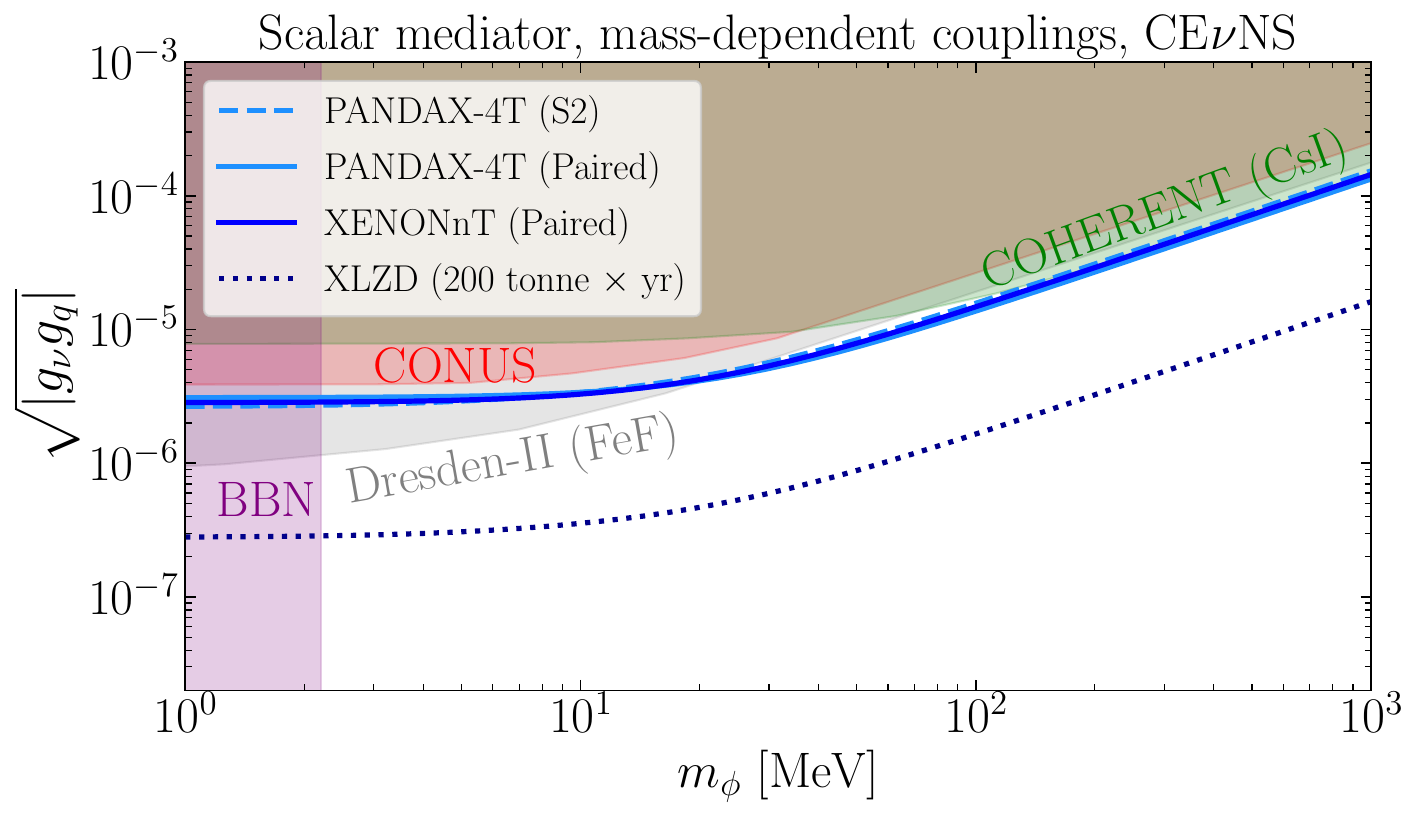} \\
        \includegraphics[width=0.48\textwidth]{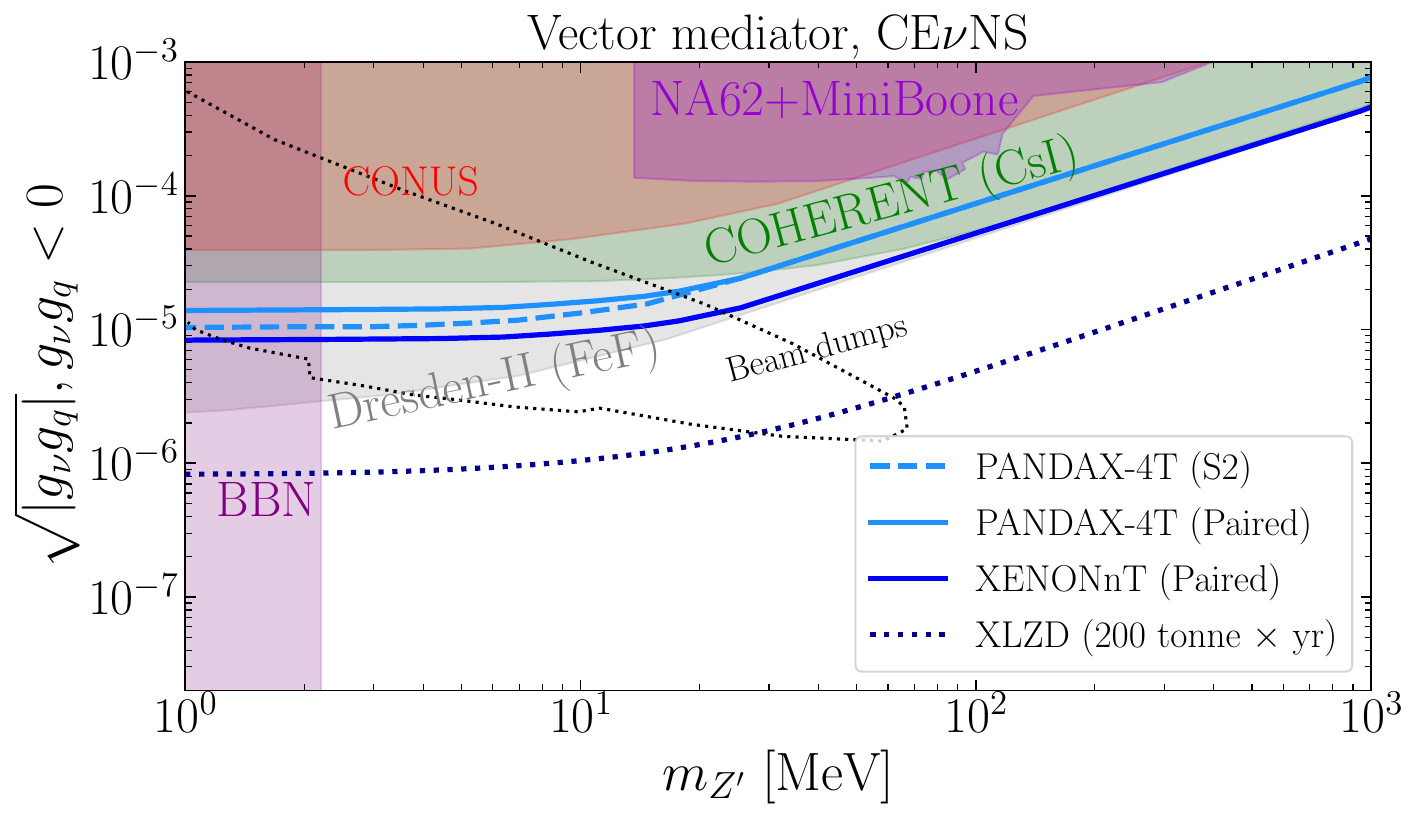} \hfill &
        \includegraphics[width=0.48\textwidth]{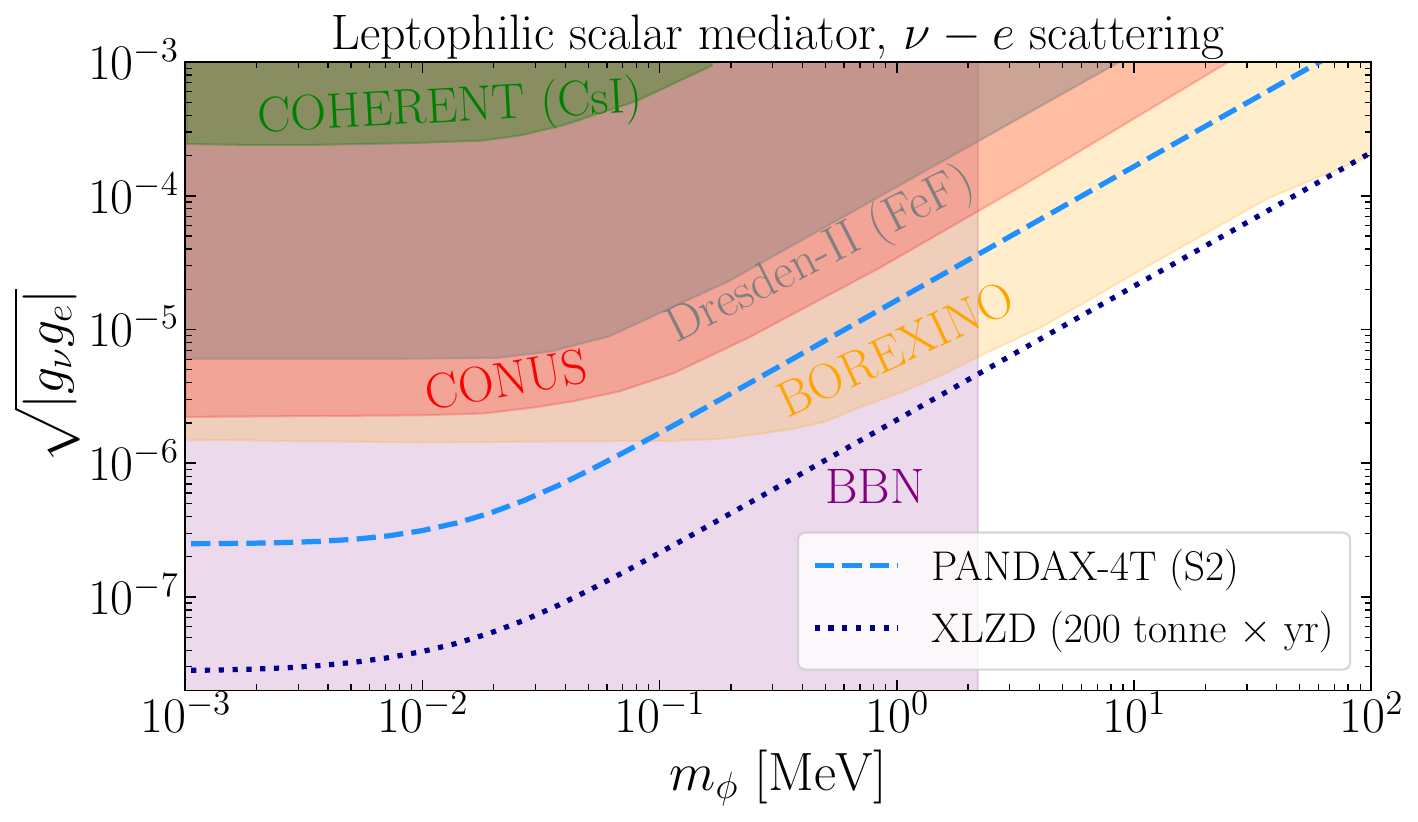}
    \end{tabular}
    \caption{\textbf{Top left:} limits on the mass and couplings of a new scalar mediator (see \cref{eq:L-scalar}), from the requirement that coherent elastic neutrino--nucleus scattering (CE$\nu$NS) mediated by this new particle shall not violate the 90\% C.L.\ constraints from PandaX-4T (light blue) and XENONnT (blue). We show results for both the S1+S2 analysis (solid) and for the S2-only analysis, but find no significant differences in sensitivity. We also show the expected sensitivity of a hypothetical detector with an exposure of 200~tonne yrs (dark blue dotted). The green region is ruled out by measurements of CE$\nu$NS in the cesium iodide (CsI) detectors of COHERENT. We further show upper limits from the CONUS \cite{CONUS:2021dwh} and Dresden-II \cite{Coloma:2022avw} experiments. The parameter region shaded in purple is excluded by Big Bang Nucleosynthesis (BBN) and the Cosmic Microwave Background (CMB) \cite{Huang:2017egl, Ghosh:2024cxi}.
    \textbf{Top right:} analogous limits on the mass and couplings of a new scalar mediator with couplings to quarks proportional to the quark masses, see \cref{eq:q-mass-dependence}.
    \textbf{Bottom left:} constraints on CE$\nu$NS mediated by a light vector with interactions given by \cref{eq:L-vector}. In this panel, our results are obtained assuming a negative product of couplings $g_qg_{\nu}$. Our results for positive couplings (when destructive interference with the SM amplitude may take place) can be found in Fig.~\ref{fig:degeneracy_vector}. Compared to scalar mediators, we further show bounds in shaded orange on a leptophobic vector from NA62 \cite{NA62:2019meo} and MiniBoone \cite{MiniBooNE:2017nqe}. Furthermore, the region within the dotted black line is excluded by a combination of beam dump experiments from visible decays of a $Z^{\prime}$ coupled to baryon number, if further assuming $g_q=g_{\nu}$. We obtained these bounds with \texttt{DarkCast} \cite{Ilten:2018crw}, and arise from a combination of $\nu$-CAL \cite{Blumlein:2013cua, Blumlein:2011mv} and NOMAD/PS191 \cite{Gninenko:2011uv} via $\pi^{0} \rightarrow Z^{\prime}\gamma$. These constraints are significantly relaxed for vector mediators which do not couple to electrons at tree-level. \textbf{Bottom right:} constraints on neutrino--electron scattering through a new leptophilic scalar mediator (see \cref{eq:L-leptophilic}) from the PandaX-4T S2-only analysis. We compare to limits from BBN, Borexino, CONUS, COHERENT (CsI), and Dresden-II \cite{Coloma:2022umy,Coloma:2022avw}.}
    \label{fig:limit}
\end{figure*}

Based on the differential scattering rates and efficiencies, together with the exposure times, background predictions, and observed number of events summarized in \cref{tab:exp-params}, we can constrain the coupling of new light mediators. We compute the ratio of the Poisson likelihoods of the observed event number in both a background-only scenario (which we define including the SM contribution to the solar neutrino event rates) and in the scenario including new physics. By Wilks' theorem \cite{Wilks:1938dza}, this likelihood ratio follows a $\chi^2$ distribution if the number of events is sufficiently large, so we can set limits based on the quantiles of the $\chi^2$ probability distribution (for 1 degree of freedom, which implies $\chi^2 < 2.71$). We find the following upper limits at 90\% confidence level on the extra contribution to the CE$\nu$NS rate due to new physics:
\begin{align}
   -2.10 < &\, \Delta N^{\rm PA,S1+S2}_{\rm sig} <  2.01~& \mathrm{(PandaX-4T, S1+S2)} \nonumber  \\
   -5.79 <  &\, \Delta N^{\rm PA,S2}_{\rm sig} \hspace{5 mm}   < 35.59~&    \mathrm{(PandaX-4T, S2~only)} \nonumber \\
   -10.50 <  &\, \Delta N^{\rm XE,S1+S2}_{\rm sig} <  9.71~&  \mathrm{(XENONnT, S1+S2)}   
\end{align}
By imposing that the event rate for a given parameter point shall not exceed these limits, we find the upper limits on the new mediator mass $m_\phi$ or $m_{Z'}$, and on the product of couplings $g_\nu g_q$ shown in \cref{fig:limit}. The four panels of this figure correspond to a scalar mediator with universal couplings to quarks according to \cref{eq:L-scalar} (top left), a scalar mediator with quark mass-dependent couplings (top right), a vector mediator according to \cref{eq:L-vector} (bottom left), and a leptophilic scalar (bottom right).  We have verified our procedure by checking that our upper limit on a new heavy vector mediator is comparable to the limit on new four-fermion interactions derived in Ref.~\cite{AristizabalSierra:2024nwf}. For this comparison, we relate the couplings $g_\nu$, $g_q$ to the dimensionless parameters $\epsilon_q$ from Ref.~\cite{AristizabalSierra:2024nwf} via $\epsilon_q = g_\nu g_q / (2 \sqrt{2} G_F m_{Z^\prime}^2)$.

The qualitative shape of the PandaX-4T and XENONnT bounds is the same in all panels of \cref{fig:limit}: they scale as $m_\phi^{-1}$ down to mediator masses $m_{\phi}\simeq 2 m_A E_{\rm er} \sim \SI{30}{MeV}$ (or $m_{\phi}\simeq 2 m_e E_{\rm er} \sim \SI{10}{keV}$ in the case of neutrino--electron scattering for the leptophilic mediator). At lower mediator masses, the curves flatten out, as can be easily understood from \cref{eq:xsec-sm-7s,eq:xsec-scalar,eq:QV-vector}. The PandaX-4T S2-only limits present a slightly different $m_\phi$ dependence compared to XENONnT and PandaX-4T S1+S2 because the PandaX-4T S2-only analysis is sensitive down to lower recoil energies (see in \cref{fig:eff}). Overall, PandaX-4T's S2-only data provides the least stringent of the limits derived here due to the smaller exposure.

\begin{figure}
    \centering
    \includegraphics[width=\columnwidth]{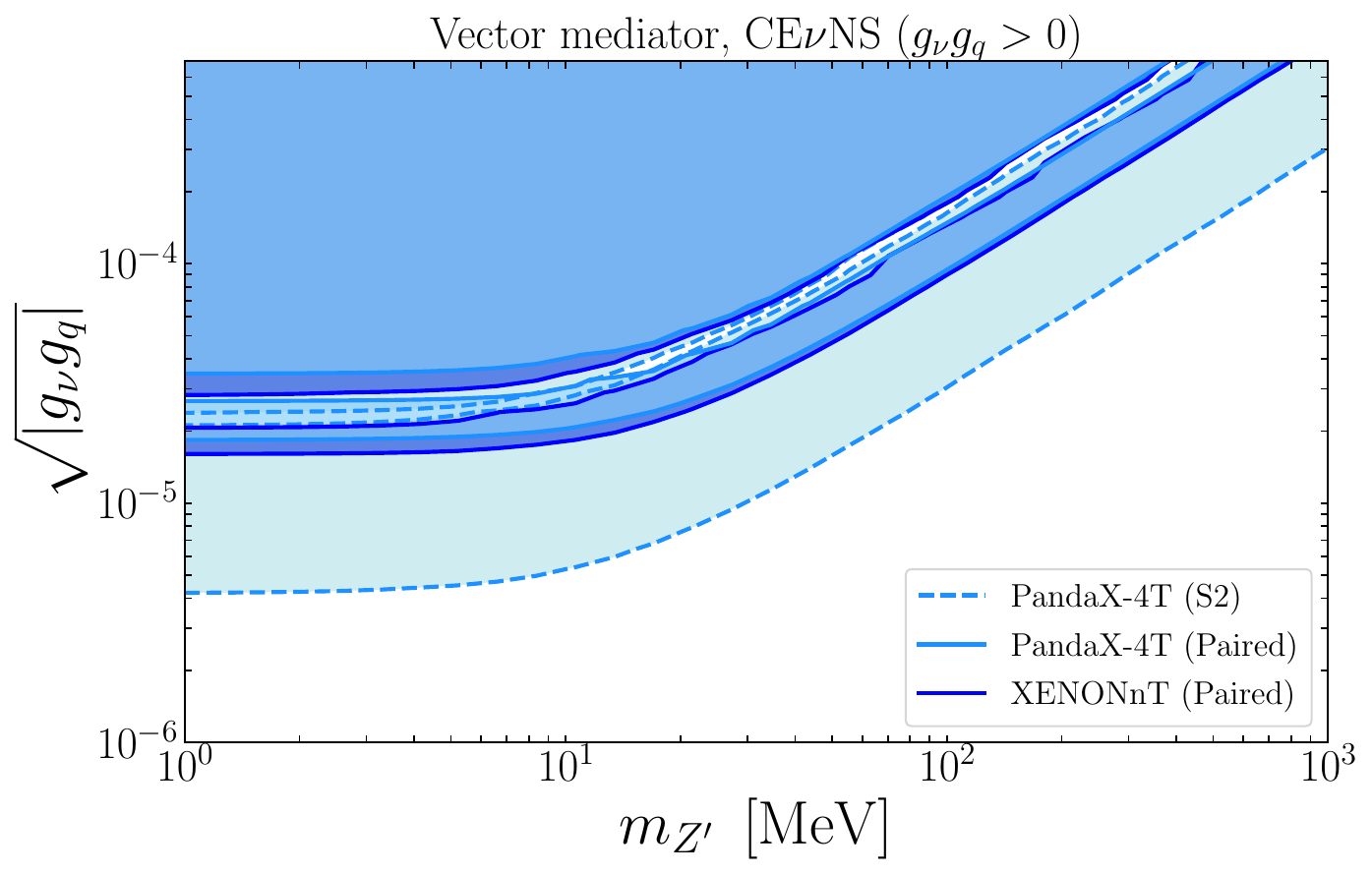}
    \caption{Limits on the couplings versus mass of a new vector mediator at PandaX-4T (S2 only, light blue, dashed; paired analysis, light blue, solid) and XENONnT (blue, solid). These limits are analogous to those shown in the lower left panel in Fig. \ref{fig:limit}, but for the case when destructive interference with the SM is allowed (that is, for $g_\nu g_q > 0$). See main text for further details.}
    \label{fig:degeneracy_vector}
\end{figure}

Comparing the top left and bottom left panels of \cref{fig:limit}, we note that the limits on vector mediators are less tight than those for flavor-universal scalar mediators by a factor of $\sim 2$. This may at first seem surprising, given that new vector interactions may interfere constructively with the SM contributions, while for new scalar interactions, there is no interference. However, the scalar charge of the nucleus is larger than the vector charge, which overcompensates for the absence of interference. Finally, we point out that for the vector mediator case there is a region of parameter space where destructive interference with the SM contribution does not allow to place a limit. We show these additional regions at PandaX-4T and XENONnT in Fig. \ref{fig:degeneracy_vector}. Note that, for each experiment separately, there is a ``band'' for intermediate values of the couplings which cannot be excluded. This is easy to understand from Eq.~\eqref{eq:QV-vector}, where we see that, if
\begin{equation}
\frac{3 g_q g_\nu}{\sqrt{2} G_F}
            \frac{(Z + N) \, F_N(E_{\rm nr})}{|q|^2 + m_{Z^{\prime}}^2} \simeq - 2Q_V \, , \label{eq:degeneracy} 
\end{equation}
the resulting cross section would match the SM result in Eq.~\eqref{eq:xsec-sm-7s}, leading to a result that is consistent with the observed data. In fact, in the limit $|q|^2 \ll m_{Z^\prime}^2$ the degeneracy with the SM expectation is exact, while the dependence with the momentum transfer breaks it for $m_{Z^\prime}^2 \lesssim |q|^2$. Since for direct detection experiments we are using only the total event rate information, the degeneracy remains unbroken and the band is visible for each experiment separately regardless of the mass of the mediator. We also note that the relative differences among the bounds obtained from the different data sets are amplified for a vector mediator compared to a scalar mediator. This is expected due to the interfering vector term scaling linearly with the product of couplings $g_qg_\nu$, while the scalar case scales quadratically $g_q^2g_\nu^2$. This brings the upper limits close to each other when plotted in terms of $\sqrt{g_{\nu}g_q}$, due to the $\mathcal{O}(1)$ differences in exposures, upper limits on the number of signal events, and experimental efficiencies which drive the sensitivity of each experiment.

We have also derived projected upper limits for the planned XENON--LUX--ZEPLIN--DARWIN (XLZD) experiment \cite{Aalbers:2022dzr}, which is expected to achieve an exposure of \SI{200}{tonne\,yrs}. We assume an energy threshold of $E^{\rm XLZD}_{\rm thr} = \SI{0.1}{keV}$, an efficiency equal to 1 in the ROI, and an upper limit on the number of signal events of $N^{90\%, \rm XLZD}_{\rm sig} < 2.71$. As expected, we find a significant enhancement of the sensitivity by about one order of magnitude with respect to current limits. Such an enhancement can become important to probe neutrinophilic dark matter models at the MeV scale \cite{Berryman:2022hds}.

%-----------------------------------------------------------------------------
\subsection{Comparison with Other Limits}
%-----------------------------------------------------------------------------

Let us now compare the limits we have derived from PandaX-4T and XENONnT data to those from other CE$\nu$NS experiments, notably COHERENT \cite{COHERENT:2017ipa,DeRomeri:2022twg}, CONUS \cite{CONUS:2021dwh} and Dresden-II \cite{Colaresi:2022obx, Coloma:2022avw}. Further details on how we obtain the COHERENT limits are given in the Appendix. For CONUS, we use a low-energy quenching factor derived from a Lindhard model with parameter $k=0.16$. For Dresden-II, we use a quenching factor determined using iron-filtered monochromatic neutrons (Fef) \cite{Coloma:2022avw}; see Ref.~\cite{AristizabalSierra:2022axl} for an analysis of Dresden-II data using the Lindhard model instead). The plots show that the dependence of the CE$\nu$NS limits from PandaX-4T and XENONnT on the mediator mass is different from the one in COHERENT \cite{DeRomeri:2022twg}. Direct detection experiments have a lower recoil energy threshold $E_{R,\rm thr}$ than COHERENT, so according to the scaling arguments given above the new physics-mediated scattering cross section scales linearly with the mediator mass down to lower $m_{\phi}$, $m_{Z'}$. Consequently, while at mediator masses $m_\phi, m_{Z'} \gtrsim \SI{50}{MeV}$, our bounds are comparable to those from COHERENT, they are stronger by a factor of $\sim 2$--4 at lower masses. Compared to Dresden-II, PandaX-4T and XENONnT CE$\nu$NS bounds are weaker at $m_{\phi}$, $m_{Z^\prime} \lesssim \SI{1}{MeV}$, but stronger at larger mediator masses. (If Dresden-II data is analyzed using a quenching factor based on the Lindhard model, the crossing point shifts from $\sim \SI{1}{MeV}$ to $\sim \SI{3}{MeV}$.)

At masses below $m_{\phi} \lesssim \SI{1}{MeV}$, new light mediators are also strongly constrained by the Big Bang Nucleosynthesis (BBN) bound on new relativistic particle species \cite{Blinov:2019gcj}.\footnote{Weaker but complementary constraints to the ones derived in Ref.~\cite{Blinov:2019gcj} could be obtained with the observation of a future Supernovae \cite{Suliga:2020jfa}.}. Finally, for the vector mediator, complementary constraints exist from beam dump experiments \cite{Blumlein:2013cua, Blumlein:2011mv, Gninenko:2011uv}, which already exclude a portion of parameter space accessible to CE$\nu$NS (see the dotted gray exclusion region in \cref{fig:limit} lower left panel). These bounds apply for a leptophobic vector which couples to neutrinos and quarks. Such a scenario can be realized, for instance, if neutrinos mix with sterile states charged under gauged baryon number \cite{Pospelov:2011ha, Harnik:2012ni}. Similar bounds from beam dump experiments and colliders may apply to scalar mediators as well, but recasting such bounds is beyond the scope of this work.

%-----------------------------------------------------------------------------
\subsection{Limits on Leptophilic Scalar Mediators}
%-----------------------------------------------------------------------------

We finally discuss the constraints on leptophilic scalar mediators shown in the bottom right panel of \cref{fig:limit}. In deriving these limits, we have included the ${}^{8}$B, CNO, ${}^{7}$ Be and $pp$ solar neutrino fluxes, where the latter is by far the dominant contribution to neutrino--electron scattering in the ROI of PandaX-4T and XENONnT. We find that solar neutrinos from ${}^{8}$B, CNO and ${}^{7}$ Be contribute less than 1$\%$ to the total electron scattering rate in the ROI of PandaX-4T. We focus here on the PandaX-4T S2-only analysis, since in the combined S1+S2 data sets, electron recoil events are efficiently subtracted. To convert the efficiency curves shown in \cref{fig:eff} from nuclear recoil to electron recoil equivalent energies, we use the Lindhard model with $k=0.133 Z^{2 / 3} A^{-1 / 2}$ \cite{Schwemberger:2022fjl}. Furthermore, neglecting BSM physics arising from CE$\nu$NS, we can use an upper limit on the number of signal events of $N^{\rm XE,S1+S2}_{\rm sig} <  9.71$. We have also derived projected limits from the future XLZD experiment, using the same experimental parameters as above.

We find that the limits from PandaX-4T (S2) improve upon those obtained from complementary laboratory experiments like BOREXINO, CONUS, Dresden-II and COHERENT at sufficiently small mediator masses $m_{\phi} \lesssim \SI{0.1}{MeV}$ thanks to the smaller energy threshold compared to BOREXINO and the larger exposure compared to CONUS. However, the existence of such light scalars would be in tension with BBN. PandaX-4T is not competitive with BOREXINO at larger mediator masses, and only a future XLZD-type detector could achieve a (marginal) improvement.

%=============================================================================
\section{Conclusions}
\label{sec:conclusions}
%=============================================================================

The recent observation of a ``fog'' or ``floor'' of solar neutrino interactions in the dark matter direct detection experiments PandaX-4T and XENONnT marks a milestone in astroparticle physics, opening a new window to possible physics beyond the SM in the neutrino sector. In the first part of the present paper, we have discussed the event rates due to CE$\nu$NS, the Migdal effect, and neutrino--electron scattering in the energy window of interest to dark matter experiments. The fact that the Migdal effect and neutrino--electron scattering are non-negligible impacts in particular the analyses based solely on the ionization signal (S2-only anlyses). (In nuclear recoil analyses, the Migdal effect is included in the quenching factor calibration, and neutrino--electron scattering events are efficiently rejected.)

We have then derived novel constraints on light scalar and vector mediators using both the S1+S2 and the S2-only datasets of PandaX-4T and XENONnT. We have found that both experiments provide leading constraints at mediator masses $\lesssim \SI{50}{MeV}$ thanks to their extremely low energy thresholds and low background rates. At larger mediator masses, their constraints are similar to those from COHERENT.

Large-scale dark matter direct detection experiments have become exquisite neutrino detectors, and a precise understanding of the event rate at these experiments is not only crucial to claim a future potential dark matter signal, but also to elucidate if neutrinos have interactions beyond the Standard Model.

%=============================================================================
\section*{Acknowledgments}
%=============================================================================

We would like to thank Daniel Pershey and Alexey Konovalov for useful discussions regarding the latest analysis of COHERENT CsI data. PBM would like to thank Pablo Muñoz Candela for useful discussions regarding the computation of the CsI weak charges. PBM would like to thank IPPP for hospitality during his visit, where part of this work was carried out. The work of GH, PH and IS is supported by the U.S. Department of Energy under award number DE-SC0020262. The work of PBM is supported by the Spanish MIU through the National Program FPU (grant number FPU22/03600). PC is supported by grant RYC2018-024240-I, funded by MCIN/AEI/10.13039/501100011033 and by “ESF Investing in your future”. The work of ZT is supported by Pitt PACC and CERN's Theoretical Physics department. This work has been partially supported by the Spanish Research Agency through grant CNS2023-145338 funded by MCIN/AEI/10.13039/501100011033 and by “European Union NextGenerationEU/PRTR”, and through Grant PID2022-142545NB-C21 funded by MCIN/AEI/10.13039/501100011033/ FEDER, UE. This project has received support from the European Union’s Horizon 2020 research and innovation programme under the Marie Skłodowska-Curie grant agreement No 860881-HIDDeN and No 101086085-ASYMMETRY, and from the Spanish Research Agency (Agencia Estatal de Investigación) through the Grant IFT Centro de Excelencia Severo Ochoa No CEX2020-001007-S funded by MCIN/AEI/10.13039/501100011033. We also acknowledge the use of the Hydra Cluster at IFT.

%=============================================================================
\section*{Note added}
%=============================================================================

During the completion of this work, Ref.~\cite{DeRomeri:2024iaw} appeared discussing the impact of light mediators on elastic neutrino--electron scattering and CE$\nu$NS at PandaX-4T and XENONnT. Our results are qualitatively comparable to these, but we also derive prospects for the future XLZD experiment. We further discuss the impact of the Migdal effect in these experiments.

%=============================================================================
\appendix
\section{Deriving Limits from the CsI Detectors in COHERENT}
\label{ap:COHERENT}
%=============================================================================

COHERENT is an experiment designed to measure coherent elastic neutrino--nucleus scattering (CE$\nu$NS) using neutrinos from a stopped-pion source. In 2017, they achieved the first observation of CE$\nu$NS \cite{COHERENT:2017ipa}, and it has since released a substantial amount of additional data \cite{COHERENT:2021xmm}. Here, we explain our procedure for fitting this data.

COHERENT present their data as a list of events, each characterized by its reconstructed energy, $E_{\rm rec}$, and time relative to the arrival of the primary proton beam spill, $t_{\rm rec}$. The expected number of events in the $i$-th energy bin and $j$-th time bin is
\begin{multline}
    N^{ij} = N_T \! \int \!\! dt \, dE_{\rm rec} \, dE_{\rm nr} \, dE_\nu \,
        \varepsilon_E(E_{\rm rec}) \, \varepsilon_t(t_{\rm rec}) \\
    \times
        \mathcal{R}(E_{\rm rec}, E_{\rm nr})
        \sum_{\substack{\alpha=\mu,\bar\mu,e \\ A={\rm Cs}, {\rm I}}} \!\!\!
        \frac{d\sigma^{\alpha A}(E_{\rm nr}, E_\nu)}{dE_{\rm nr}}
        f^\alpha(E_\nu) \, g^\alpha(t_{\rm rec}) , 
    \label{eq:CsI_events}
\end{multline}
where $E_\nu$ is the neutrino energy, $E_{\rm nr}$ is the nuclear recoil energy, $\frac{d\sigma^{\alpha A}}{dE_{\rm nr}}$ is the CE$\nu$NS cross-section for neutrinos of flavor $\alpha$ on target nuclei of species $A$ (see \cref{eq:xsec-sm-7s}), and the detector response function $\mathcal{R}(E_{\rm rec}, E_{\rm nr})$ maps nuclear recoil energies onto reconstructed energies. The functions $\varepsilon_E(E_{\rm rec})$ and $\varepsilon_t(t_{\rm rec})$ are efficiency factors. Finally, $f(E_\nu) \, g(t_{\rm rec})$ is the neutrino flux, factorized into an energy spectrum and a time profile. The function $f(E_{\nu})$ is taken from Ref.~\cite{Coloma_2020}, while the others inputs are taken from the COHERENT data release accompanying Ref.~\cite{COHERENT:2021xmm} as ancillary files. The sum in the second line of \cref{eq:CsI_events} runs over the three neutrino flavors produced in $\pi^+$ decay and over the two target isotopes. The integrals over $t_{\rm rec}$ and $E_{\rm rec}$ run over the width of bin $(ij)$, while the $E_{\rm nr}$ integral runs over the interval $(0, m_\mu/2)$, where $m_\mu$ is the muon mass. For the neutrino energy, the integration is performed over the range $(E_\nu^{\rm min}, E_\nu^{\rm max})$, where $E_\nu^{\rm min}$ is the minimum neutrino energy for a given recoil energy (see \cref{eq:E-nu-min-nr}) and $E_\nu^{\rm max}$ corresponds to the endpoint for each flavor in $\pi^+$ decay.

We consider three sources of background: a steady-state background, and two beam related backgrounds, namely beam-related neutrons and neutrino-induced backgrounds. These are described and parametrized in the supplementary materials of the data release accompanying Ref.~\cite{COHERENT:2021xmm}.

To incorporate the effect of a new light mediator we add to the cross section in \cref{eq:CsI_events} the new physics contribution according to \cref{eq:xsec-scalar,eq:xsec-leptophilic} for the scalar mediator, or according to reference \cite{Hoferichter_2020}, modifying the Wilson coefficients to reintroduce the mediator's contribution (which yields comparable results to using  \cref{eq:xsec-sm-7s} with the replacement \cref{eq:QV-vector}), for the vector mediator.

Finally, we perform a maximum-likelihood fit using a Poissonian $\chi^2$ to compare the predicted event rate to the observed one from Ref.~\cite{COHERENT:2021xmm}. We introduce six nuisance parameters that parameterize variations in the neutrino-to proton yield (essentially, the normalization for the number of events), the normalization of the three backgrounds, the energy efficiency parameters (assumed to be fully correlated), and the time at which the neutrino burst sets on. The latter nuisance parameter is left unconstrained, while the rest are assumed to be small.

\begin{figure}[h]
    \centering
    \includegraphics[width=\columnwidth]{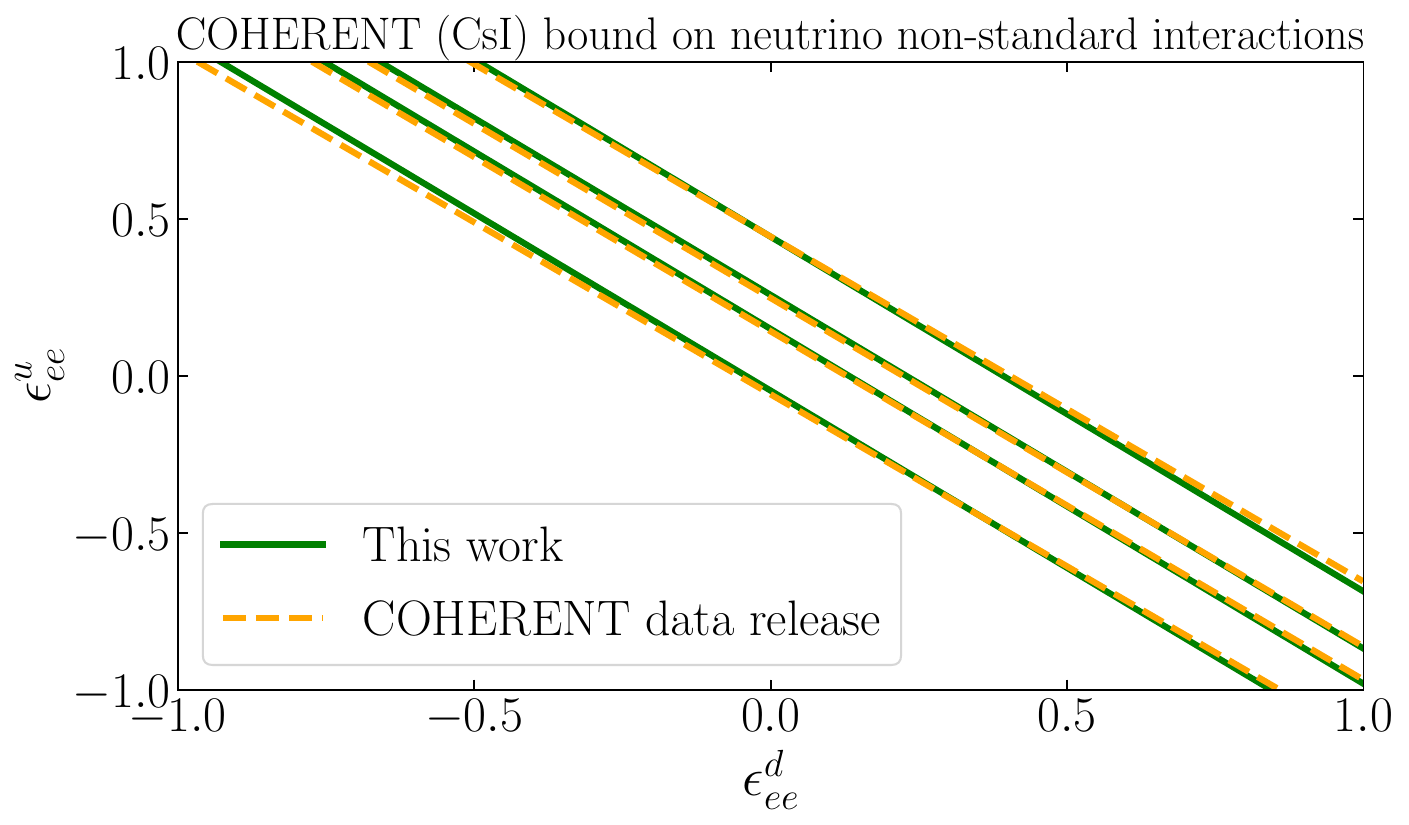}
    \caption{90$\%$ C.L.\ bounds on neutrino NSI via the parameters $\epsilon^{u}_{ee}$ and $\epsilon^{d}_{ee}$ from CE$\nu$NS at COHERENT (CsI). Our bounds (green) are compared to those from Ref.~\cite{COHERENT:2021xmm}(orange).}
    \label{fig:coherent_fit}
\end{figure}

This yields the bounds shown in \cref{fig:limit}, which are comparable (up to a factor $\sim 1.5$) to those obtained in Ref.~\cite{DeRomeri:2022twg} for CsI in both the scalar and vector cases. We further compare our results with those from the COHERENT data release in Ref.~\cite{COHERENT:2021xmm} in \cref{fig:coherent_fit}. In this plot, we show in particular bounds on neutrino--quark Non-Standard Interaction (NSI), corresponding to the limit where the mediator mass squared is much larger than the momentum transfer of the scattering. It can be seen that our bounds agree very well with those from the experimental data release.

%=============================================================================
\bibliography{References}
%=============================================================================

\end{document}